# First experiences with the Intel MIC architecture at LRZ

Volker Weinberg and Momme Allalen, LRZ

With the rapidly growing demand for computing power new accelerator based architectures have entered the world of high performance computing since around 5 years. In particular GPGPUs have recently become very popular, however programming GPGPUs using programming languages like CUDA or OpenCL is cumbersome and error-prone. Trying to overcome these difficulties, Intel developed their own Many Integrated Core (MIC) architecture which can be programmed using standard parallel programming techniques like OpenMP and MPI. In the beginning of 2013, the first production-level cards named Intel Xeon Phi came on the market. The currently fastest supercomputer in the world, Tianhe-2, heading the TOP500 list published in June 2013 uses Intel Xeon Phi coprocessors to speed up its peak performance to 34 PFlop/s. LRZ has been considered by Intel as a leading research centre for evaluating coprocessors based on the MIC architecture since 2010 under strict NDA. Since the Intel Xeon Phi is now generally available, we can share our experience on programming Intel's new MIC architecture.

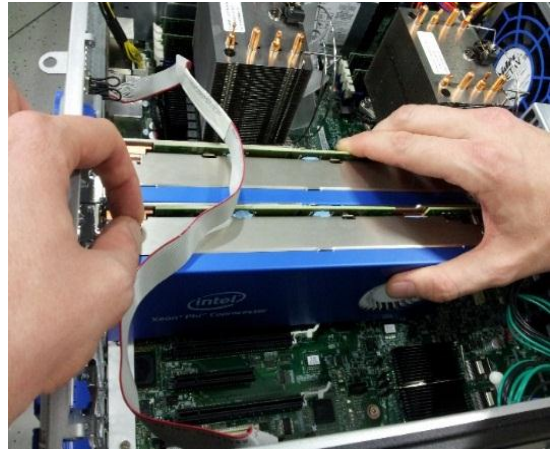

*Figure 1: Installation of 2 Intel Xeon Phi cards at LRZ.*

## Intel MIC architecture at LRZ

LRZ has already been selected by Intel in late 2010 as one of a few centres worldwide to evaluate the new Intel MIC architecture. Since then LRZ got various prototypes of this hardware, starting with a "Knights Ferry" prototype with 32 cores and 1 GB RAM and later various hardware stepping preproduction cards of the "Knights Corner" prototype with approximately 60 cores and up to 8 GB RAM. Initial results have been presented at the ISC '11 and under NDA in September 2012 at a workshop organised by IBM in Montpellier. LRZ is also part of the European DEEP project which develops a prototype for a potential exascale-enabling computer based on Intel Xeon Phi coprocessors and a special Terabit EXTOLL network. Currently within the PRACE-3IP project LRZ leads the subtask to create a Best Practice Guide for Intel Xeon Phi [1]. We plan to install a small Intel Xeon Phi cluster at LRZ in near future and currently prepare a MIC programming workshop in our course curriculum for 2014. Within PRACE, LRZ will analyse the scalability of the seismic community code SeisSol on up to 64 Intel Xeon Phi coprocessors of the EURORA cluster at CINECA (Bologna) as a showcase for a real MPI-based geophysics application.

## Architectural overview

The Intel Xeon Phi coprocessor is an advancement of the so called "Larrabee" chip which has never become a product. Larrabee was mainly targeting the computer graphics market and could only be programmed with lots of efforts under Windows. In contrast to that, Intel Xeon Phi is primarily targeting the HPC market and can be programmed using standard parallel programming techniques under Linux. One of the main advantages of the coprocessor is that the programmer can directly login on the card from the host using TCP/IP based tools like *ssh*. This allows the user to watch and control processes running on the coprocessor using tools like *top* and *kill* and gaining the benefit of all useful information available via the Linux */proc* filesystem.

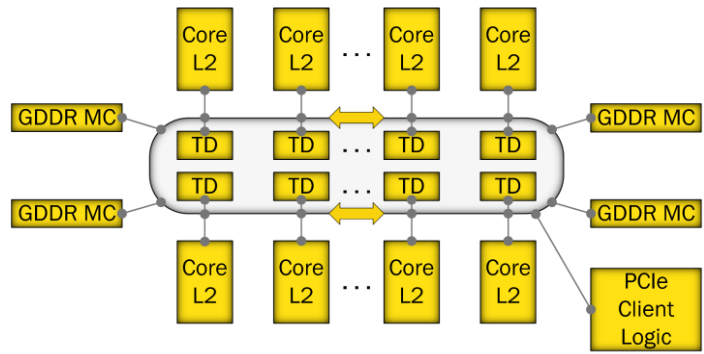

*Figure 2: Overview of the Intel MIC architecture.*

An architectural overview is given in Fig. 2. A bidirectional ring interconnect connects all the cores, L2 caches and other components like the tag directories (TD), the PCIe client logic or the GDDR5 memory controllers. Details about our currently installed Intel Xeon Phi card (C0-ES2 stepping) are summarised in the following table.

| Number of cores | 57 |
|---|---|
| Frequency of cores | 1.1 GHz |
| GDDR5 memory size | 6 GB |
| Number of hardware threads | 4 |
| SIMD vector registers | 32 (512-bit wide) per thread context |
| Flops/cycle | 16 (DP), 32 (SP) |
| Theoretical peak performance | 1 TFlop/s (DP), 2 TFlop/s (SP) |
| L2 cache per core | 512 kB |

## Programming models

The main advantage of the MIC architecture is the possibility to program the chip using plain C, C++ or Fortran and standard parallelisation models like OpenMP, MPI and hybrid OpenMP & MPI. The coprocessor can also be programmed using Intel Cilk Plus, Intel Threading Building Blocks, pthreads and OpenCL. OpenACC support is / will be provided by companies like CAPS or PGI. Standard math-libraries like Intel MKL are supported and last but not least the whole Intel tool chain, e.g. Intel C/C++ and Fortran compiler, debugger and Intel VTune Amplifier. It is also possible to do hardware-specific tuning using Intrinsics or Assembler. However, we would not recommend doing this (except maybe for some critical kernel routines), since MIC vector Intrinsics and Assembler instructions are incompatible with SSE or AVX instructions.

Generally speaking, two main execution modes can be distinguished: native mode and offload mode. In "native mode" the Intel compiler is instructed (through the use of the compiler-switch *–mmic*) to cross-compile for the MIC architecture. This is also possible for OpenMP and MPI codes. The generated executable has to be copied to the coprocessor and can be launched from within a shell running on the coprocessor.

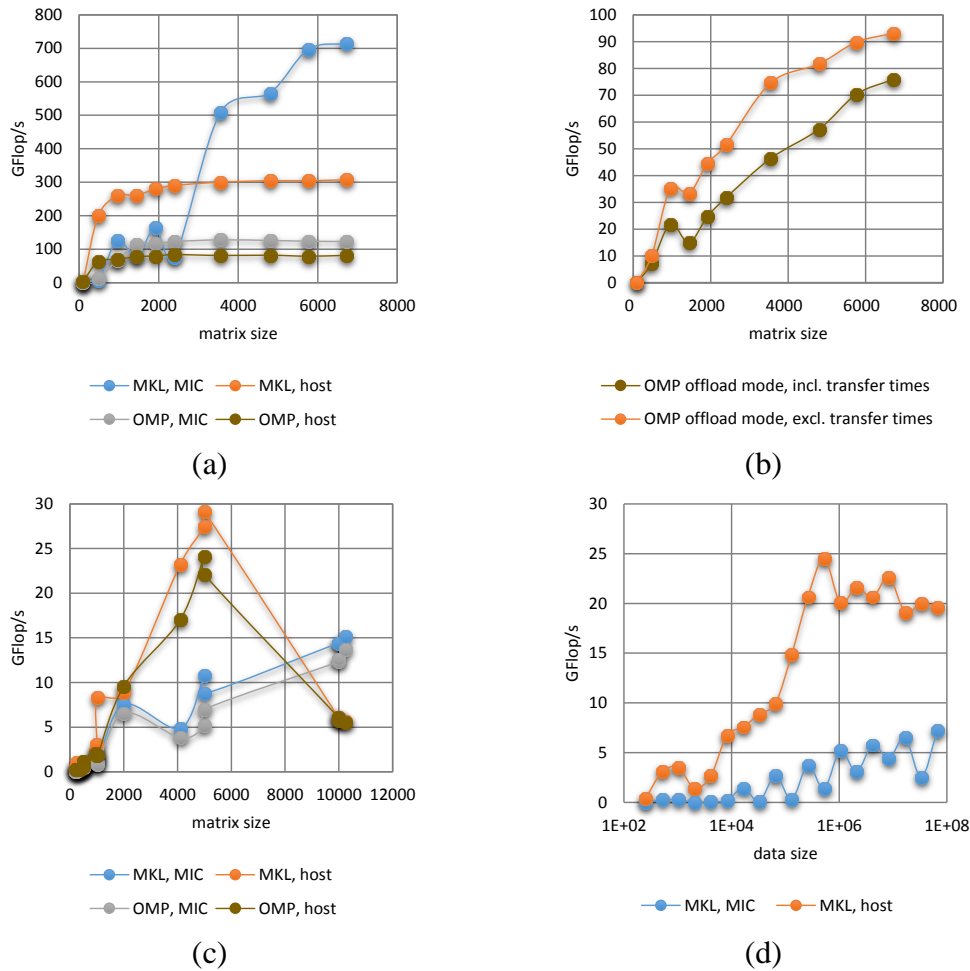

*Figure 3: Performance results (double precision) for the 3 EuroBen kernels MxM (a) & (b), SpMxV (c) and FFT (d). Fig. (a), (c) and (d) compare the performance of MKL and/or OpenMP (OMP) based implementations on Intel Xeon Phi using approx. 228 threads in native mode with the performance on the SandyBridge-EP based host system using 16 threads. Fig. (b) shows the OpenMP performance of MxM using offload mode, both including and excluding data transfer times between the host and the coprocessor.*

In "offload mode" the code is instrumented with OpenMP-like pragmas in C/C++ or comments in Fortran to mark regions of code that should be offloaded to the coprocessor and be executed there at runtime. The code in the marked regions can be multithreaded by using e.g. OpenMP. The generated executable must be launched from the host. This approach is quite similar to the accelerator pragmas introduced by the PGI compiler, CAPS HMPP or OpenACC to offload code to GPGPUs.

For MPI programs, MPI ranks can reside on only the coprocessor(s), on only the host(s) (possibly doing offloading), or on both the host(s) and the coprocessor(s) allowing various combinations in clusters.

## Benchmarks

We have ported several mathematical and kernel benchmarks partly from the SuperMUC benchmark suite to the Intel MIC architecture. For this article we concentrate on three synthetic benchmarks from the EuroBen benchmark suite [2]: a dense matrix-matrix multiplication (MxM), a sparse matrix-vector multiplication (SpMxV) and a one-dimensional complex Fast Fourier Transformation (FFT). As representatives of three (dense linear algebra, sparse linear algebra and spectral methods) of the "seven dwarfs of HPC" they have been heavily used within PRACE to analyse 12 programming languages and paradigms with respect to their performance and programmability under the leadership of LRZ [3].
Performance results for MxM (a) & (b), SpMxV (c) and FFT (d) are shown in Fig. 3. Subfigures (a), (c) and (d) compare the double precision performance on the Intel Xeon Phi coprocessor in native mode (57 cores @ 1.1 GHz with 1 TFlop/s peak) with the performance on the SandyBridge-EP based host (16 cores @ 2.6 GHz with 333 GFlop/s peak) using MKL and/or an optimised OpenMP implementation. For MxM, 714 GFlop/s could be reached using MKL on the coprocessor for the largest matrix size, which is 2.3 times faster than on the host. The optimised OpenMP implementation is 1.5 times faster for matrix sizes > 1500. Subfigure (b) displays the MxM performance in offload mode using OpenMP parallelisation in the offloaded code including and excluding data transfer times between the host and the coprocessor. For large matrix sizes data transfers become the bottleneck. In the case of SpMxV (c) the native performance on the coprocessor only excesses the host performance for the largest data sizes of the input parameter set by a factor of 2.5. The native MKL FFT performance (d) on the coprocessor is considerably below the host performance.

## Summary

Concerning the ease of use and the programmability Intel Xeon Phi is a promising hardware architecture compared to other accelerators like GPGPUs, FPGAs or former CELL processors or ClearSpeed cards. Codes using MPI, OpenMP or MKL etc. can be quickly ported. Some MKL routines have been highly optimised for the MIC. Due to the large SIMD width of 64 Bytes vectorisation is even more important for the MIC architecture than for Intel Xeon based systems. It is extremely simple to get a code running on Intel Xeon Phi, but getting performance out of the chip in most cases needs hard manual tuning of the code due to failing auto-vectorisation of the Intel compiler. Using Intrinsics with manual data prefetching and efficient register usage can considerably increase the performance - but completely destroys portability. We hope that the next product of the MIC family, announced as "Knights Landing" by Intel, with integrated on-package memory and also functioning in standalone CPU mode, will deliver much better performance in combination with future releases of the Intel compiler suite.

## Acknowledgements

Our work was financially supported by the KONWIHR project "OMI4papps" and by the PRACE-3IP project funded in part by the EU's 7th Framework Programme (FP7/2007-2013) under grant agreement no. RI-312763. We want to especially thank M. Klemm and M. Widmer (Intel Corp.) for continuous support and collaboration.